# On-chip erbium-doped lithium niobate microring lasers

Qiang Luo,[1,†] Chen Yang,[1,†] Ru Zhang,[1] Zhenzhong Hao,[1] Dahuai Zheng,[1] Hongde Liu,[1] Xuanyi Yu,[1] Feng Gao,[1] Fang Bo,[1,2,3,*] Yongfa Kong,[1,4] Guoquan Zhang,[1,2,5] and Jingjun Xu[1,2,6]

[1]MOE Key Laboratory of Weak-Light Nonlinear Photonics, TEDA Institute of Applied Physics and School of Physics, Nankai University, Tianjin 300457, China
[2]Collaborative Innovation Center of Extreme Optics, Shanxi University, Taiyuan 030006, China
[3]Collaborative Innovation Center of Light Manipulations and Applications, Shandong Normal University, Jinan 250358, China
[4]e-mail: kongyf@nankai.edu.cn
[5]e-mail: zhanggq@nankai.edu.cn
[6]e-mail: jjxu@nankai.edu.cn
*Corresponding author: bofang@nankai.edu.cn

**Lithium niobate on insulator (LNOI), regarded as an important candidate platform for optical integration due to its excellent nonlinear, electro-optic and other physical properties, has become a research hotspot. Light source, as an essential component for integrated optical system, is urgently needed. In this paper, we reported the realization of 1550-nm band on-chip LNOI microlasers based on erbium-doped LNOI ring cavities with loaded quality factors higher than one million, which were fabricated by using electron beam lithography and inductively coupled plasma reactive ion etching processes. These microlasers demonstrated a low pump threshold of ~20 µW and stable performance under the pump of a 980-nm band continuous laser. Comb-like laser spectra spanning from 1510 nm to 1580 nm were observed in high pump power regime, which lays the foundation of the realization of pulsed laser and frequency combs on rare-earth ion doped LNOI platform. This work has effectively promoted the development of on-chip integrated active LNOI devices.**

Photonic integrated circuit has attracted widespread attention of researchers since its proposal. From the view point of material system, integrated optical platforms based on silicon on insulator, silicon nitride, and indium phosphate have gradually developed into mature industrial scale. In addition, lithium niobite (LN) crystal, known as optical silicon by virtue of wide transparent window, attractive nonlinear optical properties, is an ideal platform for integrated optics. In recent years, with the commercialization of lithium niobate on insulator (LNOI) and the breakthrough of micro-fabrication technologies, LNOI high quality factor micro-disk cavities[1-4], microring cavities[5], and high refractive index contrast, low transmission loss ridge waveguides[6, 7] and other integrated photonic devices were fabricated, which shows much better performances than the devices based on Ti-diffused or proton exchanged weak waveguides on the lithium niobate bulk material. Based on the LNOI integrated devices, a series of nonlinear optical effects, such as sum frequency generation[8, 9], second harmonic generation[6, 10-17] and optical parametric oscillation[18], have been demonstrated under weak pump lower than 1 mW. Moreover, the applications of electro-optical modulators[19-21], optical frequency comb[22-26], microwave-to-optical conversion[27], spectrometer[28] were realized on the LNOI platform. LNOI is showing its tremendous potential to be a competitive integrated photonic platform[29, 30].

Generally, an integrated optical system is mainly composed of active and passive optical components. Lasers, as an indispensable part of integrated optical system, were reported very recently based on erbium-doped LNOI micro-disk resonators[31-33]. However, these micro-disk laser lack of stability and potential scalability due to coupling with tapered fibers in experiments. An effective solution is to integrate microring resonators with bus waveguides on the same chip. Compared with micro-disks, microrings have smaller mode volume resulting in higher power density, which can greatly enhance the interaction between light and matter.

In this paper, we reported the fabrication of waveguide-coupled erbium-doped LNOI microring resonators with quality ($Q$) factors up to $10^6$, based on which 1550-nm lasers with a threshold of ~20 µW and differential conversion efficiency of $6.61 \times 10^{-5}$% were realized. Furthermore, electro-optic modulator and other modulation devices could be integrated on the same chip in a scalable manner to enhance laser stability and expand laser application scenarios.

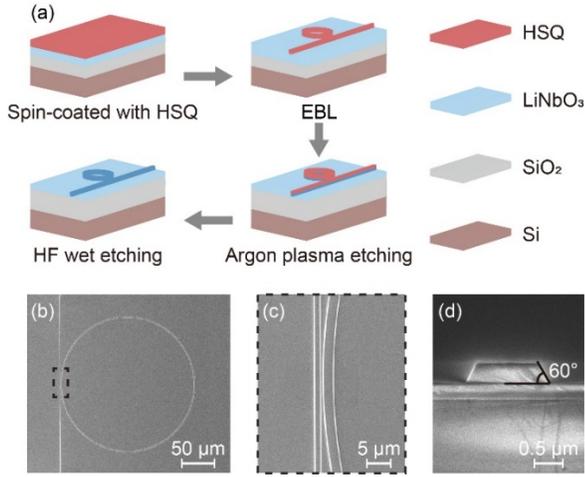

**Fig. 1.** (a) The schematic fabrication processes of erbium-doped LNOI microring resonators coupled with a waveguide. (b) Scanning electron microscope image of a waveguide-coupled erbium-doped LNOI microring resonator. (c) Zoom in of the coupling region of (b). (d) Side view of the end face of a typical waveguide.

Erbium-doped microring resonators were fabricated on an erbium-doped Z-cut LNOI wafer with a doping concentration of 0.1 mol.%. The thickness of the erbium-doped LN film, silicon-dioxide buffer layer and silicon substrate were 0.6, 2.0 and 500 μm, respectively. The fabrication processes are schematically illustrated in Fig. 1(a). Firstly, the waveguide-coupled microring patterns were defined by electron beam lithography (EBL) with hydrogen silsesquioxane (HSQ) resist. The radius and width of the microring was designed to be 100 μm and 1.8 μm, respectively, while the width of the coupling waveguide was set to be 1.0 μm. The gap between the waveguide and the ring was set as 0.65 μm to obtain efficient coupling. Subsequently, the patterns were transferred into the LN thin film using Ar$^+$ plasma etching in an inductively coupled plasma reactive ion etching (ICP-RIE) machine. The etching depth and wedge angle of the waveguides and rings are about 320 nm and 60°, respectively. Finally, the chip was immersed in buffered HF solution for 5 min to remove residual resist mask and the optical-quality end-facets were prepared by mechanical cleaving for efficient fiber-to-chip coupling. Figure 1(b) shows the scanning electron microscope image of a microring coupled with a waveguide. Magnified views of the coupling region (Fig. 1(c)) and the cross section (Fig. 1(d)) of the waveguide reveal a smooth surface and end facet.

To study the lasing action of the erbium-doped LNOI microring resonators, a 980-nm band continuous laser was employed as a pump considering the absorption coefficient of erbium ions in 980-nm band is higher than that in 1480-nm band[34]. The experimental setup to detect the light from the ring resonator is shown in Fig. 2. The pump was first divided into two parts after passing the variable optical attenuator (VOA) and coupler 1. The minor part (1%) was sent to a power meter (PM) to monitor the pump power in the optical path. The major part (99%) transmitting through a polarization controller (PC) was launched into the on-chip bus waveguide via a lensed fiber. The bus waveguide coupled light in and out of the microring resonator, inside of which erbium ions absorbed the pump and laser generated. Similarly, the light exiting the chip was collected by a second lensed fiber. The collected light was separated in to two parts through coupler 2. Light from

90%-port of coupler 2 was launched into an optical spectrum analyzer (OSA) with a response wavelength range of 600-1700 nm, to detect the 1550-nm band signal. Light from 10%-port of coupler 2 was sent to a photodetector (PD), and its output electrical signals were coupled into an oscilloscope (OSC) to monitor the transmission spectrum of the pump mode. In addition, through the external driving function, a sawtooth voltage signal generated by an arbitrary function generator (AFG) was applied on the pump laser to finely tune the pump wavelength. In the meantime, AFG provided a trigger signal to the oscilloscope, which made the transmission spectrum on the oscilloscope displayed stably.

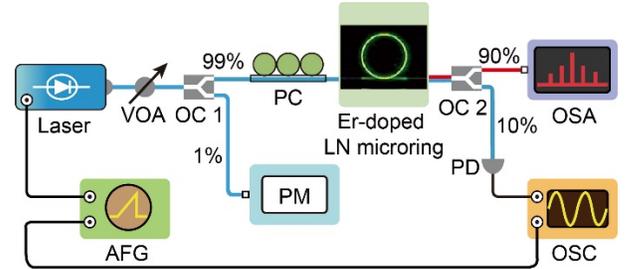

**Fig. 2.** Experimental setup for lasing action investigation in erbium-doped LNOI microring resonators. AFG: arbitrary function generator; VOA: variable optical attenuator; OC: optical coupler; PC: polarization controller; PM: power meter; PD: photodetector; OSA: optical spectrum analyzer; OSC: oscilloscope. The photograph of erbium-doped LNOI ring clearly shows the green up-conversion fluorescence.

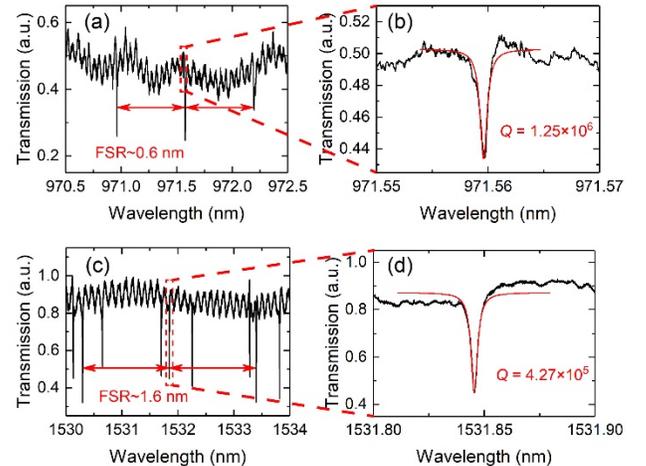

**Fig. 3.** (a) Transmission spectrum of an erbium-doped LNOI ring in 980-nm band. (b) Enlarged view of the highlighted resonance mode in (a) showing a loaded $Q$-factor of $1.25 \times 10^6$ near 974.5 nm. Red curve is the Lorentz fitting of the experimental data. (c) Transmission spectrum of an erbium-doped LNOI ring in 1550-nm band. (d) Zoom in of the marked resonance mode in (c). The Lorentz fitting (red curve) reveals a load $Q$-factor of $4.27 \times 10^5$ near 1531.8 nm.

In experiments, the coupling between two lensed fiber and on-chip bus waveguide were optimized firstly. During the experiment, the single-port coupling efficiency of pump and signal were stably at approximately 20 ± 1.5% and 16.7%, respectively. Then, the transmission spectra of the erbium-doped microring resonator were measured by using wavelength scanning method. The transmission spectrum in 980-nm band is shown in Fig. 3(a), the free spectrum range in this band was obtained to be ~0.6 nm. Through fitting the transmission spectrum with Lorentz function, the loaded *Q* factor near 974.5 nm was derived to be $1.25 \times 10^6$ as shown in Fig. 3(b). Similarly, the transmission spectrum in 1550-nm band is displayed in Fig. 3(c) and Fig. 3(d). Around 1531.8 nm, a free spectrum range of 1.6 nm and a load *Q* factor of $4.27 \times 10^5$ were obtained, respectively. It should be noted that the lower *Q* value at 1531.8 nm (near the main absorption peak of erbium ions) is mainly due to the absorption of erbium ions.

Next, we adjusted the pump wavelength in the high absorption band (970-980 nm) of erbium ions while monitoring the optimal signal in the 1550-nm band from OSA. Figure 4(a) shows the emission spectrum in the range of 1531.50-1532.65 nm at 46.4 μW pump power. We found that the optimal signal value is around 1532 nm, which locates at the strongest fluorescence peak of erbium ion in the 1550-nm band. Moreover, under high pump power (~1 mW), we observed multi-peak signals in a wide bandwidth ranging from 1510 nm to 1580 nm (as shown in Fig. 4(b)) and strong green up-conversion fluorescence as displayed on the microring photograph in Fig. 2. The multiple sets of equidistant multi-peak signals are due to the broadband gain of erbium ions in this band combined with the microring resonator for frequency selective filtering. It is expected to realize ultra-short pluses and frequency combs in the future by further optimizing the dispersion of the microring resonator and introducing additional control methods such as mode-locking.

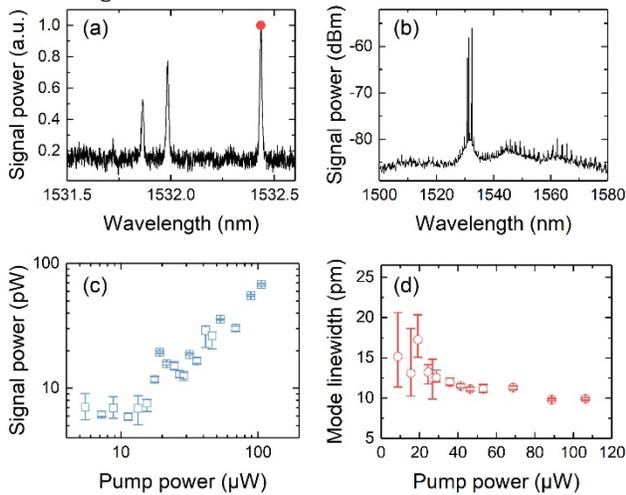

**Fig. 4.** (a) The collected emission spectrum in the range of 1531.50-1532.65 nm at 46.4 μW pump power. (b) Multi-peak lasing signal observed at a pump power of ~1 mW. (c) Power and (d) linewidth of the marked peak in (a) under different pump power.

Subsequently, with the help of the AFG, a sawtooth voltage is applied to the pump laser to tune its wavelength and thus identify the pump mode corresponding to the signal. Accordingly, the coupling depth of the pump mode and the signal power was optimized by adjusting the pump polarization. After that, we fixed the polarization state and cutoff the sawtooth signal to ensure that the pump wavelength was set at the resonance of the ring resonator. We collected the photoluminescence spectra in the range of 1531.50-1532.65 nm under different pump power, from which the power and the linewidth of the signals in 1550-nm band can be obtained. During the data collection process, the pump wavelength is first set to the red detuning side of the pump mode and is then manually adjusted to the deepest point of the pump coupling (the lowest point on the transmission line seen from the oscilloscope) to obtain the maximum signal gain under this pump power point. In this process, we used the photorefractive effect of lithium niobate for assistance[35]. Specifically, when the pump wavelength is adjusted from red detuning to blue detuning, the pump mode is broadened, and the wavelength can be fixed in the resonance, so that the intracavity energy can be effectively accumulated, which is beneficial to realize a low threshold laser.

Considering the transmission loss in the optical path and the coupling loss between the on-chip bus waveguide and the lensed fiber at both wavebands, the dependence of the power of the strongest signal marked in Fig. 4(a) by the red dot is depicted in Fig. 4(c). We can see that the signal power changes with the pump power in an S-shaped trend under the log-log scale, indicating the transition process from fluorescence emission to laser emission. In addition, we also noticed that the coupling state of the pump mode at high power begins to show unstability affected by the thermo-optical effect, so that the optimal value of signal cannot be collected. In the later research, it may be possible to achieve high power and stable laser output by use of the electro-optic effect to compensate the thermal effect and photorefractive effect of lithium niobate.

Figure 4(d) shows the relationship between the linewidth of the red-dot marked signal mode and the pump power. With the pump power increasing, the signal linewidth drops rapidly and eventually fluctuates near the resolution of the OSA (~10 pm). In fact, the linewidth of the laser should be consistent with that of the resonator mode (~1 pm). However, limited by the resolution of the spectrometer, the linewidth of the signal was imperfectly reflected. From Figs. 4 (c) and 4 (d), we can see a laser threshold of ~20 μW, at which the signal power and linewidth changed significantly with respect to the pump power. The threshold is reduced by more than one order of magnitude as compared with the previous reported LNOI microdisk lasers[31-33]. The reduction of threshold is the result of the enhancement of the pump light intensity and the spatial overlap between the pump and the signal modes by virtue of the tight confinement of microring resonators. At the same time, the differential conversion efficiency of $6.61 \times 10^{-5}$% was obtained by linearly fitting the data in Fig. 4(c). Benefiting from the integration of the bus waveguide and the microring resonator, microring lasers have much more stable performance and the potential for scalability in comparison with the tapered-fiber-coupled microdisk lasers.

In summary, we fabricated high-*Q* on-chip erbium-doped LNOI microring resonators using EBL and ICP-RIE processes. Under the pump of a 980-nm band laser, communication band lasers were demonstrated with a threshold of ~20 μW and a differential conversion efficiency of $6.61 \times 10^{-5}$%. Compared with the reported micro-disk lasers, the threshold and stability of microring lasers have been significantly improved. This work is a great advance for the development of on-chip active LNOI devices and manifests the potential of integration light sources with various functional devices based on LNOI in the future.

**Funding.** This work was supported by the National Key Research and Development Program of China (Grant No. 2019YFA0705000), the National Natural Science Foundation of China (Grant Nos. 12034010, 11734009, 92050111, 12074199, 92050114, 12004197, and 1774182), and the 111 Project (Grant No. B07013).